\documentclass[prap,showkeys,showpacs,groupedaddress,twocolumn,longbibliography]{revtex4-1}

\usepackage{graphicx}
\usepackage{dcolumn}
\usepackage{bm}
\usepackage{color}

\begin{document}
\title{Optical measurements of strong microwave fields with Rydberg atoms in a vapor cell}

\author{D.~A.~Anderson}
\author{S.~A.~Miller}
\author{G.~Raithel}
\affiliation{Department of Physics, University of Michigan, Ann Arbor, MI 48109}
\author{J.~A.~Gordon}
\author{M.~L.~Butler}
\author{C.~L.~Holloway}
\affiliation{National Institute for Standards and Technology, U.S. Department of Commerce, Boulder, CO 80305}

\date{\today}

\begin{abstract}
We present a spectral analysis of Rydberg atoms in strong microwave fields using electromagnetically induced transparency (EIT) as an all-optical readout.  The measured spectroscopic response enables optical, atom-based electric field measurements of high-power microwaves.  In our experiments, microwaves are irradiated into a room-temperature rubidium vapor cell.  The microwaves are tuned near the two-photon $65D-66D$ Rydberg transition and reach an electric field strength of 230~V/m, about 20\% of the microwave ionization threshold of these atoms.  A Floquet treatment is used to model the Rydberg level energies and their excitation rates.  We arrive at an empirical model for the field-strength distribution inside the spectroscopic cell that yields excellent overall agreement between the measured and calculated Rydberg EIT-Floquet spectra.  Using spectral features in the Floquet maps we achieve an absolute strong-field measurement precision of $6\%$.
\end{abstract}

\maketitle
\section{Introduction}
Significant progress has been made in recent years towards establishing atomic measurement standards for field quantities~\cite{Budker.2007,Sheng.2013,Sedlacek.2012}.  Rydberg atoms hold particular appeal for applications in electrometry due to their large transition electric dipole moments, which lead to a strong atomic response to electric (E) fields~\cite{Gallagher,Osterwalder.1999}.  Rydberg electromagnetically induced transparency (EIT) in atomic vapors~\cite{Mohapatra.2007} has recently been demonstrated as a practical approach to absolute measurements of radio-frequency (RF) electric (E) fields over a broad frequency range (1 to 500~GHz) suitable for the development of calibration-free broadband RF sensors~\cite{Holloway2.2014}.  The utility of the Rydberg EIT technique in characterizing RF E-fields has been demonstrated in a number of applications.  These include microwave polarization measurements~\cite{Sedlacek.2013}, millimeter-wave (mm-wave) sensing~\cite{Gordon.2014}, and sub-wavelength imaging~\cite{Holloway.2014,Fan.2014}.  The approach has also been employed in room-temperature studies of multi-photon transitions in Rydberg atoms~\cite{Anderson.2014}, as well as in measurements of static E-fields~\cite{Barredo.2013} for precise determinations of quantum defects~\cite{Grimmel.2015}.

To date, the Rydberg EIT measurement technique has been employed in measurements of weak RF fields.  In the weak-field regime, the atom-field interaction strength is small compared to the Rydberg energy-level structure, and the level shifts of the relevant coupled atom-field states are well described using perturbation theory~\cite{Dutta.2007}.  By exploiting near-resonant and resonant dipole transitions between high-lying Rydberg levels, which elicit a maximal atomic response, RF fields from as small as $\sim 1$~mV/m to a few tens of V/m have been measured~\cite{Sedlacek.2012,Anderson.2014}. For measurements of strong RF E-fields, the atom-field interaction cannot be modeled using perturbative methods, precluding the utility of the atom-based technique as demonstrated thus far for E-field measurements of high-power RF sources.  In strong fields, a new approach is necessary to accurately describe the atomic system.  Extending the atom-based measurement approach to a high power regime could enable, for example, sub-wavelength characterization of antennas radiating high-power microwaves among other applications (for an overview of high-power RF technologies see Ref.~\cite{Barker}).

In this work, we study the response of a Rydberg EIT system to strong microwave fields up to about 20\% of the microwave ionization limit of the chosen atomic Rydberg states~\cite{Gallagher,Sirko.1994}.  In this strong-field regime, the atomic response becomes highly nonlinear, and a non-perturbative Floquet treatment must be employed for an accurate model of the Rydberg level shifts and excitation rates.  The Floquet model allows us to obtain quantitative information on the E-field of the high-power RF radiation source of interest.

This paper is organized as follows.  We begin in Section~\ref{sec:2} with a description of our experimental setup.  To illustrate the contrast between the weak-field and strong-field regimes in a Rydberg EIT experiment, in Section~\ref{sec:3} we compare Rydberg EIT spectra for weak mm-waves, tuned to the one-photon $26D_{5/2}-27P_{3/2}$ Rydberg transition, with spectra for strong microwaves, tuned near the two-photon $65D-66D$ Rydberg transition.
In Section~\ref{sec:4} we outline the Floquet theory required for an accurate model of the atomic response in the strong-field regime. The theory is then employed in Section~\ref{sec:5} to extract quantitative information on the microwave field and its distribution within the spectroscopic cell used for the strong-field sample spectra in Section~\ref{sec:3}.  We conclude with a discussion on how our results may be used in atomic high-power RF measurement methods.

\section{Experimental setup}
\label{sec:2}
\begin{figure}[h]
\includegraphics[width=8.5cm]{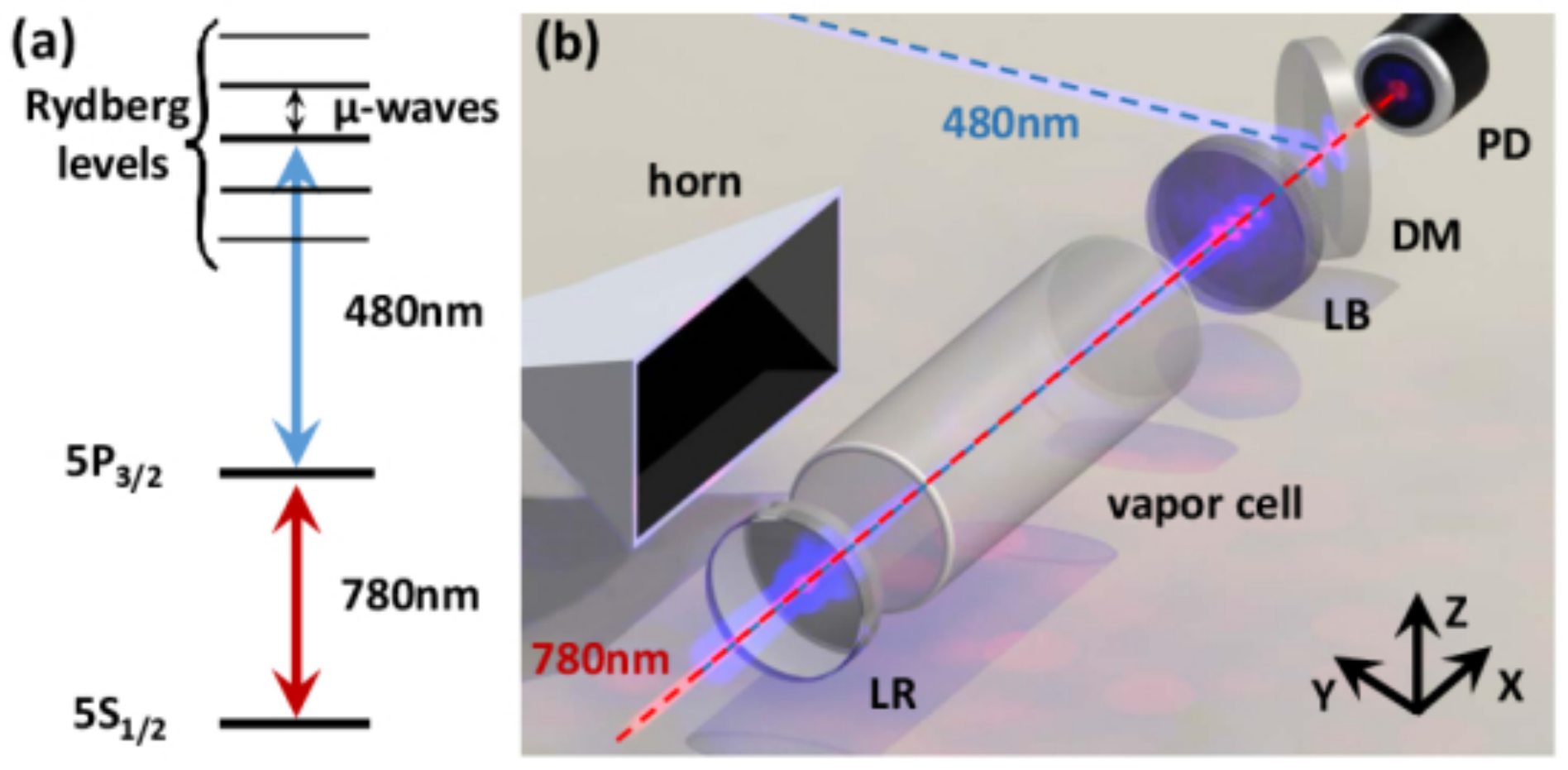}
\caption{(a) Energy-level diagram for Rydberg EIT and the Rydberg states coupled by RF fields.  (b) Illustration of the experimental setup showing the vapor cell, 480~nm and 780~nm laser beams (dashed lines), photodiode (PD), blue lens (LB), red lens (LR), dichroic mirror (DM), and microwave horn.}
\label{fig:1}
\end{figure}

The Rydberg EIT three-level system in $^{85}$Rb relevant to this work is illustrated in Fig~\ref{fig:1}(a).  It includes a weak probe laser resonant with the $5S_{1/2}(F=3)\to5P_{3/2}$ transition and a strong coupling laser resonant with the $5P_{3/2}\to$~Rydberg transition.  Application of the coupling laser results in an increased transmission of the probe field when a $5S_{1/2}\to 5P_{3/2}\to$~Rydberg double resonance condition is met.  In our experiment, the probe laser with $\lambda_p=780.24$~nm has a power of 100~$\mu$W and is focused to a beam width of 94~$\mu$m FWHM; the coupling laser with $\lambda_c\approx480$~nm has a power of 37~mW and is focused to a beam width of 144~$\mu$m FWHM.

The experimental setup is similar to one described in previous work~\cite{Anderson.2014}; a schematic is shown in Fig.~\ref{fig:1}(b).  The probe and coupling beams are counter-propagating and overlapped through a room-temperature isotopic $^{85}$Rb vapor cell and have parallel linear polarizations along $z$.  The probe laser frequency is scanned over the Doppler-broadened $5S_{1/2}\to5P_{3/2}$ transition at a rate of about 1~Hz, while the coupling laser frequency is fixed near the $5P_{3/2}\to$~Rydberg transition.  The coupling laser power is modulated by a 30~kHz square pulse with a 50\% duty cycle, while the transmission of the probe beam through the cell is detected on a photodiode.  The photodiode signal is processed with a lock-in amplifier and recorded on an oscilloscope.

Microwaves are produced by a signal generator and emitted from a horn aimed at the vapor cell perpendicular to the propagation direction of the optical beams, as shown in Fig.~\ref{fig:1}(b).  The dipole moments and frequencies of dominant RF transitions between Rydberg levels typically scale as $\sim n^2$ and $\sim n^{-3}$, respectively ($n$ is the Rydberg atom principal quantum number). Weak-field atomic response can be observed over a wide RF frequency range by matching the RF frequency with that of a strong Rydberg transition, at suitable quantum numbers, and by keeping the RF field low enough that the Rabi frequency of the Rydberg transition remains much smaller than the RF frequency and smaller than typical level separations in the unperturbed spectrum. Nonlinear response and high-order state mixing occurs when using high $n$-values, strong RF fields, or a combination of both, leading to considerably richer atom-field interaction phenomena.  In Section~\ref{sec:3} we show two sample spectra to highlight the qualitatively different nature of the atomic response in weak- and strong-field regimes. The weak-field measurement is performed in the far-field of a mm-wave horn in the WR-6 band (injection power $-18.7$ to $2.4$~dBm).  The emitted field has a linear polarization that is parallel to the optical-beam polarizations. For the strong-field measurement, microwaves in the K$_u$-band are emitted from a horn placed to within about 1~cm of the optical beam foci, which is in the near-field of the horn (injection power 13 to 24~dBm).

\section{Weak-field and strong-field measurements}
\label{sec:3}
\begin{figure}[h]
\includegraphics[width=8.7cm]{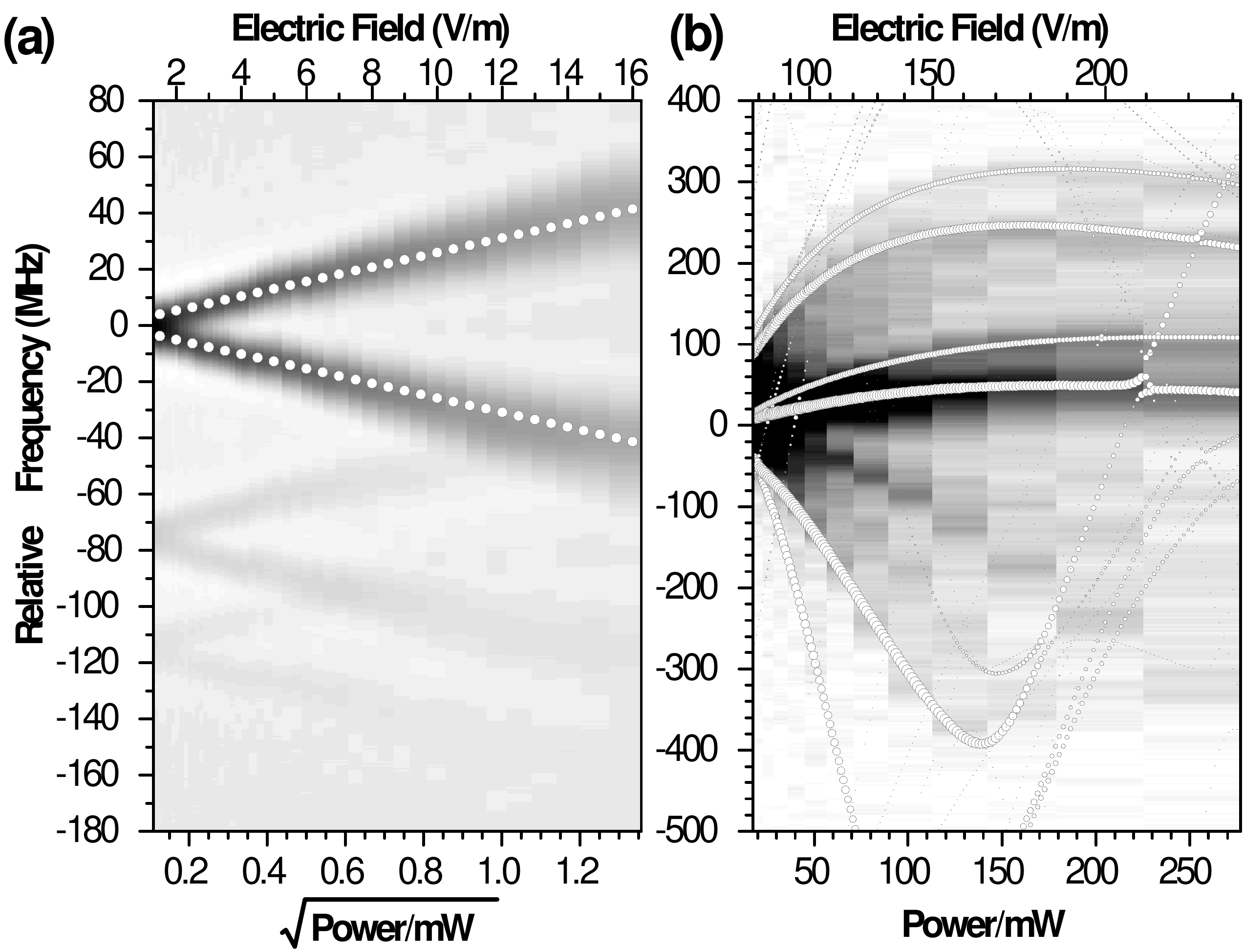}
\caption{Experimental spectra: (a) Weak-field measurement of 132.6495~GHz mm-waves on the $26D_{5/2}-27P_{3/2}$ one-photon transition versus $\sqrt{\rm power}$.  Each spectrum is an average of 25 traces and the signal is represented on a linear gray scale in arbitrary units.  A calculated spectrum is overlaid with relative excitation rates from $5P_{3/2}$, given by the dot areas, and E-field values on the top axis.  (b) Strong-field measurement of 12.4611548~GHz microwaves on the $65D-66D$ two-photon transition versus power.  Each spectrum is an average of 20 and the signal is represented on a linear gray scale.  The Floquet calculation is overlaid with excitation rates from $5P_{3/2}$, given by the dot areas, and E-field values on the top axis.  Background signal and additional features in the experimental strong-field spectrum are due to the microwave E-field inhomogeneity within the cell (see Section~\ref{sec:5}).
}
\label{fig:2}
\end{figure}

In weak RF fields, the Rydberg levels are dynamically (AC) Stark-shifted and, in the case of near- or on-resonant drive of a Rydberg transition, exhibit Autler-Townes splittings~\cite{Dutta.2007}.
For single-photon transitions, the RF E-field strength is obtained directly from the Autler-Townes splitting of the Rydberg EIT line, which is given by the Rabi frequency $\Omega={\bf d}\cdot {\bf E}/\hbar$, where ${\bf d}$ is the Rydberg transition dipole moment and $\bf{E}$ is the RF radiation E-field vector.  Figure~\ref{fig:2}(a) shows experimental spectra for the on-resonant
one-photon $26D_{5/2}-27P_{3/2}$ mm-wave (132.6495~GHz) transition as a function of the square root of mm-wave power~\footnote{The relative probe frequency scales displayed in the figures equal the measured frequency scales multiplied by a factor $\lambda_p/\lambda_c$ to account for Doppler shifts of the probe and coupling laser frequencies.}.  Here, the EIT coupling-laser frequency is resonant with the mm-wave-free $^{85}$Rb~$5P_{3/2}(F'=4)$ to $26D_{5/2}$ transition, where $F'$ denotes the intermediate-state hyperfine component.
%The frequency difference observed between the strong Autler-Townes peaks in Fig.~\ref{fig:2}(a) equals $\Omega$
As expected in weak RF fields, $\Omega$ is a linear function of the square root of power (which is proportional to $\rm{E}$).  The faint level pairs centered at about -70 and -110~MHz correspond to spectra associated with the intermediate $5P_{3/2}(F'=2,3)$ hyperfine components~\footnote{Since the Doppler factor for intermediate hyperfine splittings is $(1-\lambda_c/\lambda_p)^{-1}$, to read the correct hyperfine state splittings from Fig.~\ref{fig:2}(a) the frequency axis must be altered by a factor $\lambda_c/(\lambda_p-\lambda_c)=1.6$.}. From the measured spitting, using $\Omega={d_z}{E_z}/\hbar$ and the known values of the dipole moments for a $z$-polarized field $d_z$
($405~e a_0$ for magnetic quantum number $m_j=1/2$ and $331~e a_0$ for $m_j=3/2$) the E-field is obtained from the EIT spectra.  It is found that the maximum field reached in Fig.~\ref{fig:2}(a) is 16~V/m, about 0.02\% of the microwave ionization field of these atoms and well within the weak-field limit.

To study the atomic response in strong fields, we drive Rydberg atoms at the zero-field $65D_{5/2}-66D_{5/2}$ two-photon resonance frequency (12.4611548~GHz).  This two-photon Rydberg transition was chosen to accommodate high-power microwaves in the $K_u$ band.  Figure~\ref{fig:2}(b) shows experimental EIT spectra centered on the $65D$ level for microwave powers ranging from 13 to 24~dBm in steps of 1~dBm.  The twelve data sets are plotted as a function of power.  For this high-power measurement, the microwave-induced shifts are in the range of several hundred MHz, {\sl{i.e.}} about a factor of ten larger than the shifts of the low-power measurement in Fig.~\ref{fig:2}(a).  As detailed in Section~\ref{sec:5}, the maximum field reached in Fig.~\ref{fig:2}(b) is 230~V/m, about 20\% of the microwave field-ionization limit for that case.  We note that the field-free fine structure of the $65D$ and $66D$ states ($\sim 40$~MHz) is broken up in the strong-field regime because it is small compared to the microwave-induced shifts.  This aspect heralds the greater complexity of high-power Rydberg EIT spectra compared to low-power spectra, evident from a qualitative comparison of Figs.~\ref{fig:2}(a) and (b).

At the lowest microwave power in Fig.~\ref{fig:2}(b), the microwave interaction broadens the $65D$ EIT resonance to a FWHM width of $2\pi\times 50\pm 1$~MHz, which is a factor of 2 larger than that of the microwave-free EIT resonance (not shown).  For increasing microwave power, the EIT signal splits into multiple distinguishable spectral lines.  For two-photon transitions $\Omega\sim E^2$ and the lines are expected to shift linearly as a function of power.  Most levels in Fig.~\ref{fig:2}(b) exhibit linear shifts up to microwave powers of $\sim 70$~mW.

In strong fields, higher-order couplings lead to a redistribution of oscillator strength between many field-perturbed Rydberg states, resulting in smaller signal strengths compared to those in weak fields.  This is reflected in Fig.~\ref{fig:2}(b), where we observe a rapid initial decrease in signal strength; over the first 30~mW increase in microwave power the peak signal strengths of the individual spectral lines are reduced by more than an order of magnitude.  As the microwave power is increased further, the shifts of the spectral lines become non-linear in power, reflecting substantial state-mixing and higher-order couplings.  The transition from linear to non-linear behavior occurs gradually as a function of power and depends on the level. As seen by close inspection of Fig.~\ref{fig:2}(b), even at the lowest powers, most levels exhibit some degree of non-linearity.  A quantitative model of the complex level structure in the strong-field regime is described in detail in Section~\ref{sec:4}.

Inhomogeneous fields within the measurement volume contribute to the background and additional spectral lines observed in Fig.~\ref{fig:2}(b).  The field inhomogeneity is attributed in part to the presence of the dielectric cell, as was observed in previous work~\cite{Anderson.2014,Holloway.2014}, and to the fact that the measurement is done in the near-field of the microwave horn.  The effects of the field inhomogeneity are discussed in detail in Section~\ref{sec:5}.

\section{Floquet analysis}
\label{sec:4}
In strong fields, where typical Rabi frequencies approach or exceed atomic transition frequencies, high-order couplings become significant and perturbative approaches are no longer valid.  To model the strong-field experimental spectra we use a non-purturbative Floquet method.  Following the Floquet theorem, the solutions to Schr\"{o}dinger's equation for a time-periodic Hamiltonian $\hat{H}(t)=\hat{H}(t+T)$, where $T$ is the period of the RF field, are of the form
\begin{equation}
\Psi_{\nu}(t)={\rm{e}}^{-iW_{\nu}t/\hbar}\psi_{\nu}(t).
\end{equation}
Here, $\psi_{\nu}(t)=\psi_{\nu}(t+T)$ are the periodic Floquet modes and $W_{\nu}$ their quasienergies, with an arbitrary mode label $\nu$.  For the atom-field interaction strength of interest here, the Floquet modes can be represented using standard basis states $\vert n,\ell,j,m_j\rangle=\vert k\rangle$, {\sl {i.e.}}
\begin{equation}
\Psi_{\nu}(t)= {\rm{e}}^{-iW_{\nu}t/\hbar} \sum_k C_{\nu,k}(t)\vert k\rangle,
\end{equation}
with time-periodic (complex) coefficient functions $C_{\nu,k}(t)=C_{\nu,k}(t+T)$.  The Floquet energies $W_{\nu}$ and states $\Psi_{\nu}(t=0)$ are determined by finding the eigenvalues and vectors of the time-evolution operator $\hat{U}(t,T+t)$. The coefficient functions $C_{\nu,k}(t)$ are then obtained by integrating $\Psi_{\nu}(t)$ over one field period, $ t \in [0,T]$.

\begin{figure}[h]
\includegraphics[width=8.5cm]{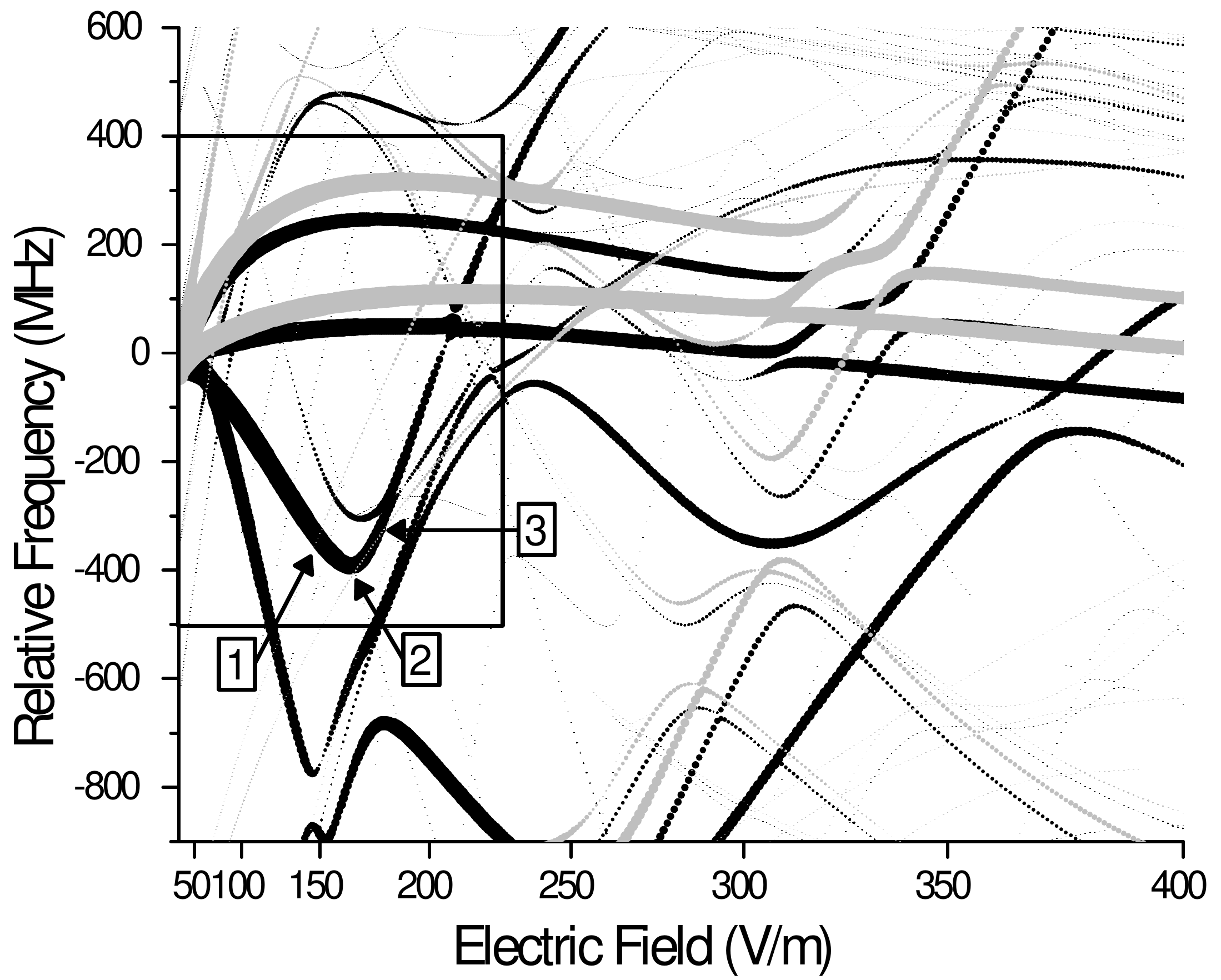}
\caption{Calculated $\vert m_j\vert=1/2$ (black) and $3/2$ (gray) Floquet quasienergies and their relative excitation rates (dot area) from $5P_{3/2}$.  The E-field axis is scaled such that linear distance is proportional to intensity, allowing a comparison with the experimental map in Fig.~\ref{fig:4}.  The region inside the black box corresponds to the parameter space covered in the experiment.  The Floquet wave-packet dynamics for the levels marked 1, 2, and 3 are shown in Fig.~\ref{fig:4}.  These calculations use a basis of all states with effective principal quantum number $56\le n^*\le73$ and orbital quantum number $\ell\le 10$.  The Floquet series are terminated at $N=\pm 8$ photon numbers.}
\label{fig:3}
\end{figure}

\begin{figure}[h]
\includegraphics[width=6.5cm]{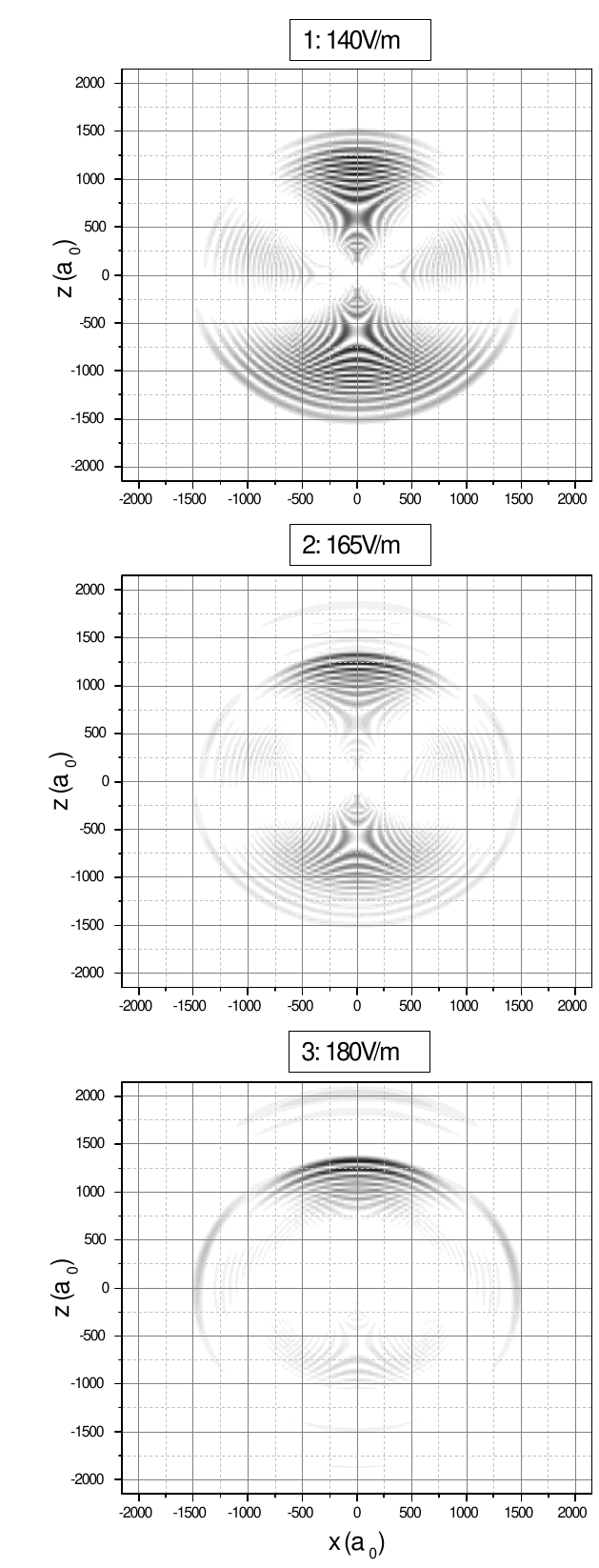}
\caption{Videos 1, 2, and 3 showing the Floquet wave-packet dynamics through one full microwave cycle for the three levels marked 1, 2, and 3, respectively, in Fig.~\ref{fig:3}.  The initial frame of each (shown) is taken at the time when the microwave E-field is maximal. The E-field points along $+z$.}
\label{fig:4}
\end{figure}

In the laser excitation of Floquet states from the intermediate $5P_{3/2}$ state, multi-photon processes are important because the atom may emit or absorb a number of microwave photons together with an optical photon. To compute excitation line strengths, the functions $C_{\nu,k}(t)$ are Fourier-expanded,
\begin{eqnarray}
\Psi_{\nu}(t)&=&{\rm{e}}^{-iW_{\nu}t/\hbar}\sum_{k}\sum^{\infty}_{N=-\infty} \tilde{C}_{\nu,k,N}{\rm{e}}^{-iN\omega_{RF}t}\vert k\rangle,\nonumber \\
\tilde{C}_{\nu,k,N}&=&\frac{1}{T}\int_{0}^{T} C_{\nu,k}(t){\rm{e}}^{iN\omega_{RF}t}dt. \quad
\end{eqnarray}
The integer $N$ is interpreted as a number of microwave photons with frequency $\omega_{RF}$ associated with the bare atomic state $\vert k\rangle$.  The laser frequencies $\omega_{L}$, where Floquet levels are resonantly excited from the $5P_{3/2}$ level, and the corresponding line strengths $S_{\nu, N}$ are then given by
\begin{eqnarray}
\hbar \omega_{L} & = & W_\nu + N \hbar \omega_{RF} \nonumber  \\
S_{\nu, N} & = & (e {E_L}/ \hbar)^2 \left| \sum_k \tilde{C}_{\nu,k,N} \, {\hat{\bf{\epsilon}}} \cdot \langle k \vert \hat{\bf{r}} \vert 5P_{3/2}, m_j \rangle \right|^2, \quad
\label{eq:rates}
\end{eqnarray}
where $E_L$ is the amplitude of the laser E-field, ${\hat{\bf{\epsilon}}}$ is the laser polarization vector, and $\langle k \vert \hat{\bf{r}} \vert 5P_{3/2}, m_j \rangle$ are the electric-dipole matrix elements of the basis states with $\vert 5P_{3/2}, m_j \rangle$.  Each Floquet level $W_\nu$ leads to multiple resonances, which are associated with the microwave photon number $N$.  This procedure is similar to one used by Yoshida {\it et al.}~\cite{Yoshida.2012}. Due to parity, in the absence of additional static fields a Floquet level $W_\nu$ may generate resonances either for even $N$ or odd $N$, but not both.

In Fig.~\ref{fig:3} we show calculated Floquet energies and excitation rates $S_{\nu, N}$ in the vicinity of the $65D$ Rydberg level for a microwave frequency of 12.4611548~GHz and field strengths ranging from 0 to 350~V/m. The field is displayed on a quadratic scale to show the dependence of the atomic level shifts on power, and for direct comparison with Fig.~\ref{fig:2}(b). Over the short frequency range displayed in Fig.~\ref{fig:3} it is always $N=0$. Further inspection of the calculated Floquet energies and excitation rates, not presented in detail here, shows that for fields above $\sim 150$~V/m several Floquet levels $W_\nu$ visible in Fig.~\ref{fig:3} have copies with high excitation rates for even $N$-values between about -8 and +8.

The Floquet modes in strong fields exhibit non-trivial wave-packet motion, and their optical excitation rates have to be calculated according to Eq.~(\ref{eq:rates}).  It would, for instance, be incorrect to associate the excitation rates of the Floquet modes in Figs.~\ref{fig:3} and~\ref{fig:5} with the $65D$ probabilities the modes carry.  To qualitatively explain this, we note that in weak fields the dressed-state coefficients and dipole moments have fixed amplitudes and phases relative to the field (in the field picture and using the rotating wave approximation).  In strong fields, wave function coefficients and dipole moments vary significantly throughout the microwave-field cycle.  Specifically, the Floquet modes are time-periodic wave-packets that are synchronized with the driving RF field.  To visualize typical Floquet wave-packet dynamics, see the videos of the wave-packet evolution of the wave functions shown in Fig.~\ref{fig:4}.  We have also investigated the wave-packet distribution over the quantum numbers $n$ and $\ell$ of the basis states. The distributions over $n$ and $\ell$ oscillate synchronously with the driving RF field, as expected, and have significant populations over the quantum-number ranges $60 \lesssim n \lesssim 73$ and $0 \leq \ell \lesssim 5$. The wave-packets therefore cover a basis-state energy range equivalent to the absorption or emission of up to about ten RF photons.

\section{Analysis of experimental spectra}
\label{sec:5}
\begin{figure}[h]
\includegraphics[width=8.5cm]{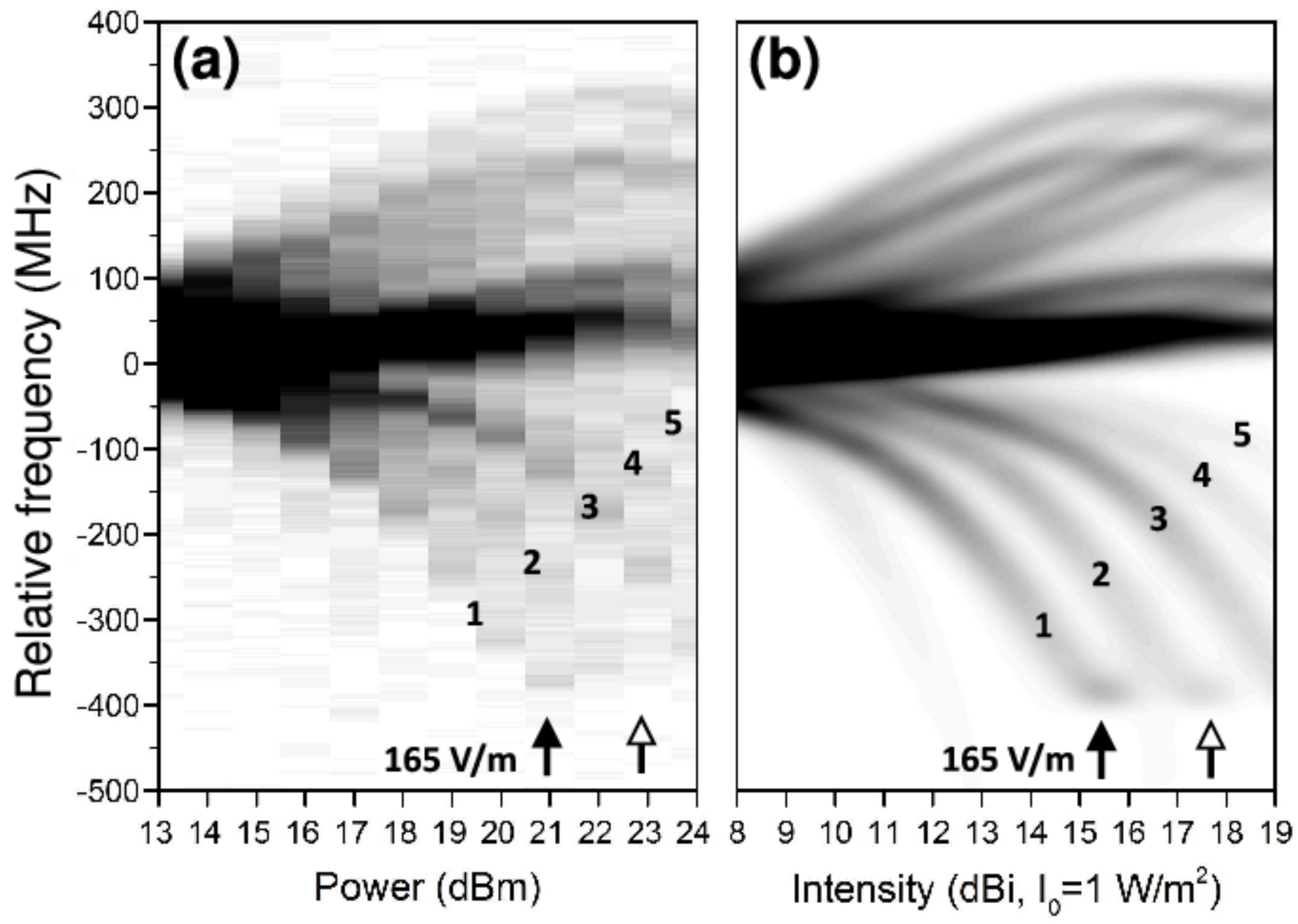}
\caption{(a) Experimental spectra of the $65D-66D$ two-photon transition versus microwave power.  The signal is represented on a linear gray scale in arbitrary units.  The arrows indicate coordinates that both correspond to the same avoided crossing in the theoretical Floquet map, marked 2 in Fig.~\ref{fig:3}. The solid arrow corresponds to the spatial region with the highest microwave intensity along the probe beam. The labels 1-5 mark the renderings of the same level in Fig.~\ref{fig:3}, observed in the five different intensity domains (see text). (b) Composite Floquet map model of (a).}
\label{fig:5}
\end{figure}

In Fig.~\ref{fig:2}(b) we find an excellent overall agreement between dominant features in the experimental and theoretical Floquet maps.  Additional features evident in the experimental map are due to E-field inhomogeneities, which are discussed in detail below.  Both experimental and theoretical maps exhibit several arched lines at positive frequencies and, at negative frequencies, several lines that shift approximately linearly in power. The dominant down-shifting line suddenly terminates at close to $-400$~MHz.  It is evident from Fig.~\ref{fig:3} that the sudden termination of the down-shifting lines is due to a wide Floquet avoided crossing.  Avoided crossings in Floquet maps, such as those in Fig.~\ref{fig:3}, provide convenient markers for spectroscopic determination of the RF E-field on an absolute scale. In the present case, the prominent avoided crossing at 165~V/m (see label 2 in Fig.~\ref{fig:3}) has several matching locations in the experimental spectrum shown in Fig.~\ref{fig:2}(b).  These are seen more clearly in Fig.~\ref{fig:5}(a), which shows the experimental data on a dBm scale.  The appearance of multiple copies of the calculated avoided crossing in the experiment points to the fact that the microwave field within the cell must have multiple dominant domains, each of which produces its own rendering of the avoided crossing.  The rendering at the lowest injected microwave power corresponds to the E-field domain with the highest field at a given injected power (calculation and E-field axis shown in Fig.~\ref{fig:2}(b)). In Figure~\ref{fig:2}(b) the avoided crossing is observed first at 130~mW, at which point the domain that has the highest field reaches 165~V/m. It follows that at a microwave power of 250~mW a maximum RF E-field of $230\pm 14$~V/m is reached.  The uncertainty is given by half the experimental step size ($\pm$0.5~dB, corresponding to $\pm 6\%$ in field).

The experimental spectrum plotted in Figs.~\ref{fig:2}(b) and ~\ref{fig:5}(a) contains information on the continuous distribution of the microwave E-field strength along the length of the EIT probe and coupling beams passing through the spectroscopic cell.  The microwave boundary conditions are symmetric about the $xy$-plane, with the incident microwave E-field being $z$-polarized and the optical beams propagating along the $x$ axis (see Fig.~\ref{fig:1}(b)).  Since the microwave field is primarily $z$-polarized along the optical beams, it drives $\Delta m_j = 0$ Rydberg transitions within the $m_j=\pm 1/2$ and $m_j=\pm 3/2$ manifolds of states.  The field distribution is a result of superpositions of reflections from the cell walls.  Also, the cell is placed within the near field of the source, leading to additional variability of the microwave field along the probe beam.  Hence, one may picture the RF field as a speckle pattern, akin to speckle patters seen for general, non-ideal coherent fields.
Here, one should expect the number of speckles to be on the order of the cell length divided by the wavelength, which in our case is on the order of three.  Also, since there are no structures within the cell that are very close to the optical probe beam path, we do not expect any sharp spatial features in the microwave E-field (features which might otherwise arise from sharp metallic or dielectric edges and the like). Therefore, for each theoretical line in Fig.~\ref{fig:3} the measured EIT spectra are expected to exhibit a small number of spectral features that correspond to the local maxima and minima of the microwave field along the length of the probe beam.

Based on the observation of five downward spectral lines [labeled in Fig.~\ref{fig:5}(a)], and the fact that the calculated spectrum has only one corresponding downward line (feature 1 in Fig.~\ref{fig:3}), we model the spectrum considering populations of atoms located within a set of five dominant microwave E-field regions.  In our model the probability distribution for intensity on a dB scale is given by
\begin{equation}
P_{dBi}(s) = \sum_{\vert m_j \vert=1/2}^{3/2} \sum_{k=1}^5 w_{m_j}(\vert m_j \vert) w_k(k)  P_{dBi0}(s+\Delta s_k).
\end{equation}
Here, $k$ is an index for the five microwave field domains, $w_{k}(k)$ is the probability that an atom contributing to the signal resides within domain $k$, $w_{m_j}(\vert m_j \vert)$ is the probability that an atom contributing to the signal has a magnetic quantum number $\vert m_j \vert$, and $P_{dBi0}$ is a Gaussian point-spread function that accounts for inhomogeneous spectral broadening within the five domains. The values $\Delta s_k$ indicate by what amount (in dB) the central microwave intensity within the $k$-th microwave field region is shifted relative to the intensity in the highest-intensity ($k=1$) domain. For $P_{dBi0}$ we assume a Gaussian that is the same for all $k$. The fit parameters in the model are $\Delta s_k$, $w_{m_j}$, $w_{k}$, and the standard deviation, $\sigma_{dBi}$, for $P_{dBi0}$.  We account for the optical EIT line broadening and laser line drifts by a Gaussian spread function in frequency, $P_{\nu}(\nu)$, which has a standard deviation $\sigma_{\nu}$. From the theoretical spectrum, $S_T(s,\nu)$, the model spectrum, $S_E(s_0,\nu_0)$, is then given by the convolution
\begin{equation}
S_E(s_0,\nu_0) = \int S_T(s_0-s', \nu_0  - \nu') P_{\nu}(\nu') P_{dBi}(s') d\nu' ds' \quad ,
\end{equation}
where the intensities in the arguments of $S_E$ and $S_T$ are measured in dBi, defined as $10 \log_{10} [I/(W/m^2)]$, where $I$ is the RF field intensity.

Figures~\ref{fig:5}(b) shows the model spectrum $S_E(s_0,\nu_0)$ in dBi for the data in Fig.~\ref{fig:5}(a).  In the model spectrum, the field domains have empirically fitted intensity shifts of $\Delta s_k = 0.0, -2.0, -4.17, -6.0, -8.0$~dB for $k=1,...,5$.  The corresponding fitted weighting factors are $w_k = 0.39, 0.21, 0.27, 0.09, 0.04$. The intensity shifts are significant to better than about 0.5~dB, while the weighting factors are significant to better than about $\pm 0.04$. The weighting factors $w_{m_j}$ for the $\vert m_j \vert=1/2$ and $3/2$ states are 0.7 and 0.3, respectively (significance level better than about 0.1). The larger $\vert m_j\vert=1/2$-weight likely results from optical pumping by the EIT probe field.  Further, $\sigma_{dBi}=1$~dBi and $\sigma_{\nu}=30$~MHz.

A comparison of measured spectrum and model spectrum in Fig.~\ref{fig:4} shows that a strong-field Floquet analysis of the atomic physics of Rydberg atoms in microwave fields, combined with a straightforward empirical model of the
microwave intensity distribution and the $\vert m_j \vert$-weighting in the sample, leads to remarkably good agreement between spectra with rather complex features.

\section{Discussion and Conclusion}

Utilizing a combination of resonant, strong electric-dipole transitions as in Fig.~\ref{fig:2}(a) and higher-order transitions such as the two-photon transition in 2(b), it is possible to observe level shifts in Rydberg EIT spectra over a wide dynamic range of the applied RF intensity. Since the spectroscopic response is well described by the Floquet theory laid out in Section~\ref{sec:4}, the measured spectra can be used to directly determine the RF E-field causing the observed spectral features in a calibration-free manner.
Specifically, there are no antenna systems and readout instruments that need to be calibrated to
translate a reading into a field, because spectral features such as line shifts and avoided crossings follow from the invariable nature of the underlying atomic physics.  The field measurement precision is given by how well the spectral features are resolved. For instance, in the present work the avoided crossing pointed out in Figs.~\ref{fig:2}(b), \ref{fig:3} and~\ref{fig:4} can be resolved with $\pm 0.5$~dBi uncertainty, corresponding to an absolute field uncertainty of $\pm6\%$.

In the weak-field domain, an example of which is seen in Fig.~\ref{fig:2}(a), the E-field strength is obtained by a measurement of the Autler-Townes splitting of the Rydberg EIT line.
In the strong-field domain, an example of which is seen in Fig.~\ref{fig:2}(b) and analyzed in depth through most of this paper, a comparison of experimental Rydberg EIT spectra with calculated Floquet maps allows us to determine the E-field strength.  Our analysis of the spectrum also reveals information about the E-field distribution within the active measurement volume.

The extension of the Rydberg EIT method to higher fields, explored in this work, paves the way to extend the dynamic range of the Rydberg EIT RF-field sensing methodology into the realm of strong microwave intensities. With the present measurement we have reached a field of 230~V/m, corresponding to a six-fold increase in field and a 40-fold increase in intensity relative to a previous measurement~\cite{Anderson.2014}. An extension of the measurement range to even higher fields will be readily achievable using
more powerful RF testing sources. Fundamental limitations due to the underlying atomic physics of Rydberg EIT will eventually occur when the Rydberg atoms experience substantial microwave ionization rates~\cite{Gallagher,Sirko.1994}.

This work was supported by the AFOSR (FA9550-10-1-0453) and DARPA's QuASAR program.
%\bibliographystyle{apsrev4-1}
%\bibliography{bibfile}

\end{document}